# Resonant Tunneling in Natural Photosynthetic Systems

Kit M. Gerodias[a], M. Victoria Carpio-Bernido, [b,c,1] and Christopher C. Bernido[b,c]

[a] *Physics Department, McGill University, 3600 rue University, Montréal QC, Canada H3A 2T8*
[b] *Physics Department, University of San Carlos, Talamban, Cebu City 6000, Philippines*
[c] *Research Center for Theoretical Physics, Central Visayan Institute Foundation, Jagna, Bohol 6308, Philippines*

**Abstract**

The high internal quantum efficiency observed in higher plants remains an outstanding problem in understanding photosynthesis. Several approaches such as quantum entanglement and quantum coherence have been explored. However, none has yet drawn an analogy between superlattices and the geometrical structure of granal thylakoids in leaves. In this paper, we calculate the transmission coefficients and perform numerical simulations using the parameters relevant to a stack of thylakoid discs. We then show that quantum resonant tunneling can occur at low effective mass of particles for 680 nm and 700 nm incident wavelengths corresponding to energies at which photosynthesis occurs.

## 1. Introduction

Photosynthesis allows organisms such as algae, plants and certain bacteria, to convert solar energy into chemical energy. The process can be divided into two reactions, light-dependent and light-independent reactions. The chemistry of photosynthetic process seems well accepted [1, 2]; however, for the physics of the light-dependent process many questions remain unresolved. In particular, the internal quantum efficiency or photon-energy-charge conversion efficiency of photosynthesis is known to be high and under certain conditions, may reach 100% [3, 4]. This amazing feat of nature is looked up to as an inspiration for the design, development, and upscaling of cost-effective technologies such as solar cells that would efficiently capture and store solar energy [5]. In contrast, for instance, Shockley and Quisser determined theoretically that a single-junction solar cell has the maximum energy conversion efficiency of only 33% [1, 3, 6]. More recently, Asahi, Kita, and their colleagues reported a structure capable of theoretically achieving up to 63% energy conversion [7]. Hitherto, the conversion efficiency in photosynthesis is not yet fully explained thus posing a major impediment for utilizing the mechanism as a viable pathway for high-efficiency solar cells [5].

---

[1] Corresponding author: mvcbernido@gmail.com

The physics of photosynthesis has been previously approached as a problem of quantum entanglement and quantum coherence [3, 8, 9]. This is not surprising since quantum entanglement was observed in the light-harvesting complexes involving proteins found in photosynthetic organisms [9]. However, calculated efficiencies still do not match nature's standards. In this paper, we provide an alternative understanding of photosynthesis by focusing on the geometric structures of thylakoids which are starkly reminiscent of semiconductor superlattices. The fact that both thylakoids and superlattices have dimensions in the order of $\sim 10^{-9}$m further compels us to draw an analogy between the two. By applying the same physics of the semiconductor nanostructure on this biological structure, we explore the behaviour and investigate the possibility of resonant tunneling as a contributing factor to the quantum efficiency of photosynthesis. Resonant tunnelling was shown by Tsu and Esaki in the early 1970's for a system with several periodic potential barriers [10], and this phenomenon has been studied and observed in different quantum systems with similar structures [11-15]. Here, we adopt the resonant tunneling approach of Yamamoto et al in their 1989 paper [16]. In Section 2, we discuss the parameters and typical dimensions of stacked thylakoids. This is followed in Section 3 by the mathematical framework for resonant tunneling. The Results and Discussion are in Section 4, and Section 5 gives the Conclusion.

## 2. Experimental Parameters in Photosynthesis

As measured through electron tomography, nanostructures inside the chloroplast are observed to have grana stacks composed of 5 to 20 thylakoids made up mostly of photosystem II (PSII) and light-harvesting complex II (LHCII) [2, 17-19]. Photosynthesis occurs at the membrane of each thylakoid where a supply of energized electrons is needed for the light-independent part of the process. The supplier of electrons is water oxidized in the thylakoid membrane through photolysis in the oxygen-evolving complex (OEC) embedded in PSII. Light is needed in PSII (P680 or pigment absorbing light at 680nm) where chlorophyll stores light energy not only to energize electrons, but also to replenish electrons by oxidizing water. Excited electrons from PSII are carried to photosystem I (PSI), called P700, via an electron transport chain within each thylakoid membrane. The electrons in PSI are further energized through a pigment absorbing light at 700nm. Hence, both photosystems PSI and PSII would need light to excite electrons [2].

In a stacked granum of, for example, 20 thylakoids [19], it is quite possible that some thylakoid membranes are not directly accessed by light. For thylakoid membranes not accessed by light, one would wonder where the required excited electrons could be obtained within each membrane. Traditional understanding says excited electrons cannot be produced in the absence of light as may happen in a thylakoid buried deep in a stacked thylakoid granum. The need for excited electrons, however, can be addressed if photons from the uppermost thylakoid accessed by light tunnels through the 20 or so stacked thylakoid disks without losing energy. This way, resonant tunneling in stacked thylakoids could shed light on the internal quantum efficiency problem.

To investigate resonant tunnelling in a stacked thylakoid granum, we note that the thickness of grana disks is typically 8.5 nm and stromal gap is 7.2 nm (see, Figure 1). The lumenal space is about 4.5 nm thick, the stromal space is 3.2 nm and the membrane is 4.0 nm [17-19]. We take the energy to be that of the incident photon, 680 nm and 700 nm [1, 2], and the membranes are assumed to act like a dielectric with capacitance from which we can calculate the barrier potential.

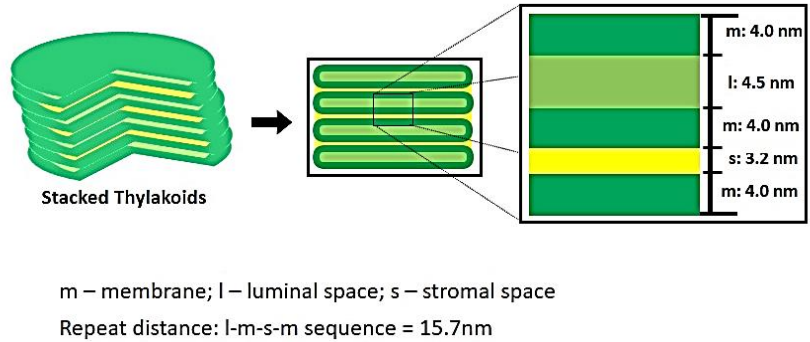

m – membrane; l – luminal space; s – stromal space
Repeat distance: l-m-s-m sequence = 15.7nm

**Figure 1.** Three adjacent membranes, each of thickness 4.0 nm, from two thylakoids separated by a 3.2 nm stromal space with distances of the intervening spaces.

## 3. Mathematical Framework

We treat the thylakoids stacked in the chloroplast as a cylindrical quantum system with periodic barrier potential along the cylindrical axis (see, Figure 2) which may be described by the Krönig-Penney model [20, 21]. In modeling the stacked thylakoids, we proceed by setting up the Schrödinger equation in circular cylindrical coordinates with $n$-barrier potentials and then resolve the radial, azimuthal and the axial parts. Since we are interested in resonant tunneling, we focus on calculating the transmission coefficient for the system up to $n$-barrier potentials. Application of the boundary conditions to the axial part, allows the transmission coefficient to be calculated from the resulting transfer matrix equations following the work of Yamamoto, et al. [16].

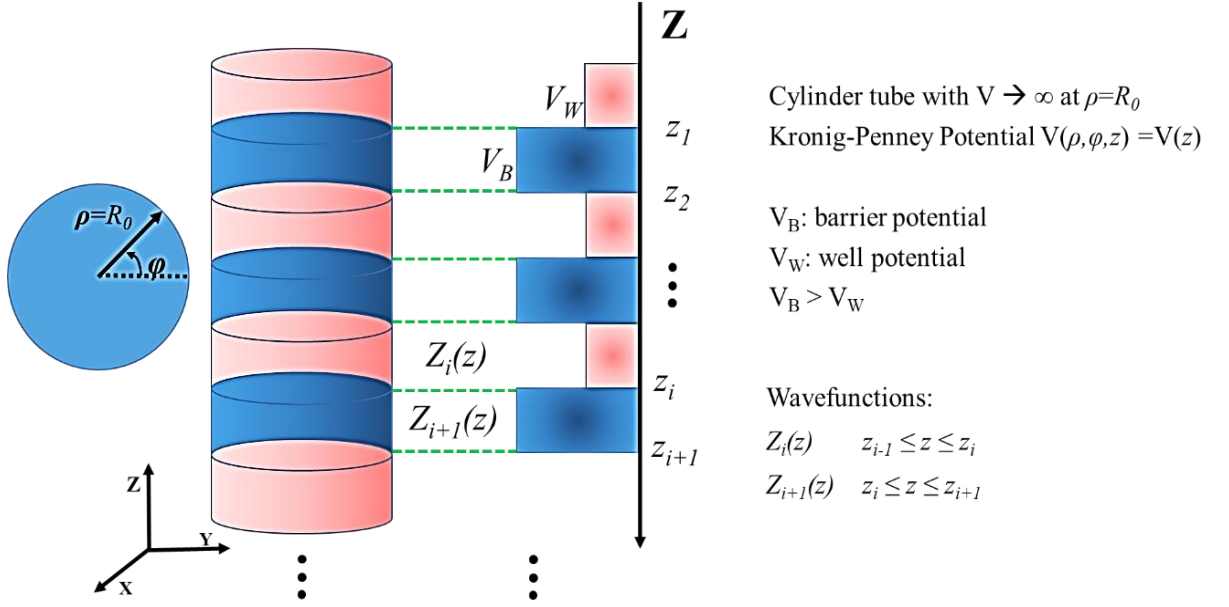

**Figure 2.** Quantum system with periodic barrier potential in cylindrical coordinates.

We write the time-independent Schrödinger equation, $(-\hbar^2/2\mu)\nabla^2\Psi + V\Psi = E\Psi$, in circular cylindrical coordinates where, $\nabla^2 = (\partial^2/\partial\rho^2) + (\partial/\rho\,\partial\rho) + (\partial^2/\rho^2\partial\varphi^2) + (\partial^2/\partial z^2)$, with, $\rho, \varphi$ and $z$ as the radial, azimuthal and axial components, respectively. Here, the potential of the system is taken as, $V(\rho, \varphi, z) = V(z) = V$ which depends only on the axial part. By separation of variables, we have the wave function, $\Psi(\rho, \varphi, z) = R(\rho)\Phi(\varphi)Z(z) = R\Phi Z$, where,

$$-\frac{\hbar^2}{2\mu}\frac{\partial^2 Z}{\partial z^2} + V_{eff}Z = EZ. \tag{1}$$

In Eq. (1), we define the effective potential $V_{eff} = \hbar^2 q^2/2\mu + V$, with the constant $q$ absorbing both radial and azimuthal parts. The solution of Eq. (1) is of the form, $\psi(z) = Ae^{-ikz} + Be^{ikz}$.

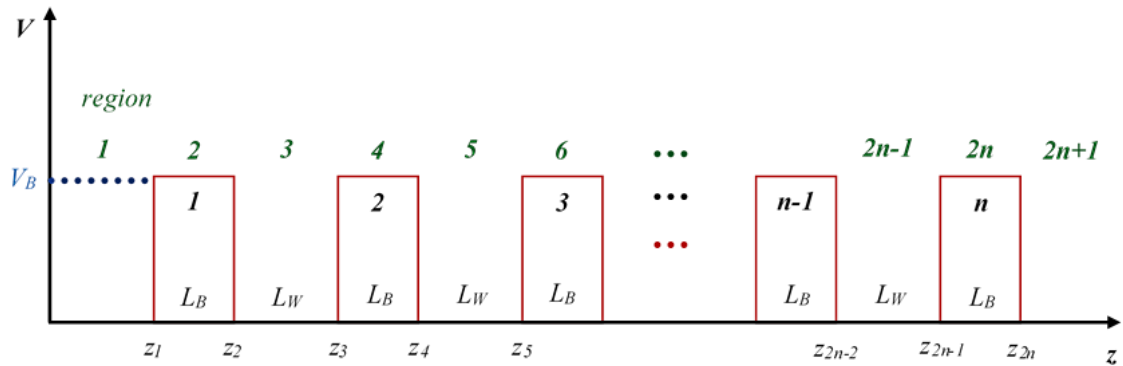

**Figure 3.** Superlattice with n-fold barrier potential where the horizontal axis is the axial part of the system.

Figure 3 shows the same superlattice as Figure 2 containing $n$ number of identical rectangular barrier potentials with thickness $L_B$ and potential height $V_B$ and $n-1$ wells of thickness $L_W$ arranged periodically in $2n + 1$ regions. From Eq. (1), the wave function at the $j$-th region is $Z_j(z) = A_j e^{ik_j z} + B_j e^{-ik_j z}$. By applying the following boundary conditions: (a) $Z_j(z_j) = Z_{j+1}(z_j)$ and, (b) $d(Z_j(z_j))/dz = d(Z_{j+1}(z_j))/dz$, we obtain the following transfer matrix equation

$$\begin{pmatrix} A_1 \\ B_1 \end{pmatrix} = R_1 R_2 R_3 \cdots R_{2n} \begin{pmatrix} A_{2n+1} \\ B_{2n+1} \end{pmatrix} \qquad (2)$$

where $R_j = \frac{1}{2k_j} \begin{pmatrix} (k_j + k_{j+1})e^{-i(k_j - k_{j+1})z_j} & (k_i - k_{i+1})e^{-i(k_j + k_{j+1})z_j} \\ (k_j - k_{j+1})e^{i(k_j + k_{j+1})z_j} & (k_j + k_{j+1})e^{i(k_j - k_{j+1})z_j} \end{pmatrix}$ and $j = 1, 2, 3, \ldots, 2n$.

By taking the following characteristic values,

$$z_2 - z_1 = z_4 - z_3 = \cdots = z_{2n} - z_{2n-1} = L_B \qquad (3a)$$

$$z_1 = z_3 - z_2 = z_5 - z_4 = \cdots = z_{2n-1} - z_{2n-2} = L_W \qquad (3b)$$

$$V_1 = V_3 = \cdots = V_{2n+1} = 0 \qquad (3c)$$

$$V_2 = V_4 = \cdots = V_{2n} = V_B \qquad (3d)$$

$$k_1 = k_3 = \cdots = k_{2n+1} = \sqrt{2\mu_W E}/\hbar \qquad (3e)$$

$$k_2 = k_4 = \cdots = k_{2n} = i\beta = i\sqrt{2\mu_B(V_B - E)}/\hbar \qquad (3f)$$

we obtain the expression for transmission coefficient $T_n$ [16] for $n$-barriers with $0 < E < V_B$,

$$T_n = (1 + A^2 Y_n^2)^{-1} \qquad (4)$$

where $A = V_B \sinh \beta L_B / 2[E(V_B - E)]^{1/2}$, with, $\beta = \sqrt{2\mu_B(V_B - E)}/\hbar$. The $Y_n$ is expressed as

$$Y_n = \sum_{i=1}^{m} -(-1)^i C(n-i, i-1) H^{n+1-2i} \qquad (5)$$

where $H = 2\cosh(\beta L_B)\cos(kL_W) - (2E - V_B)\sinh(\beta L_B)\sin(kL_W)/\sqrt{[E(V_B - E)]}$, $C(n - i, i - 1) = C(N, r) = N!/r!(N - r)!$, for index $m$, when $n$ is even, $m = n/2$ but when $n$ is odd, $m = (n+1)/2$. Therefore, resonance ($T_n = 1$) occurs when $Y_n = 0$. For any system of $n$ barriers, we can compute the expression of $Y_n$ assuming effective masses $\mu_B$ and $\mu_W$ are both equal.

From Eq. (4), the transmission coefficient $T_n$ is dependent on the dynamics of $A^2 Y_n^2$. If this term is zero, obviously, we have a transmission equal to unity, $T_n = 1$, meaning full or resonant transmission. Since, for a different number of barriers, the expression of $Y_n$ changes, we need to look at this term and check conditions at which $Y_n = 0$. From Eq. (5), we can see that $Y_n = Y(H)$.

Note that $H$, and ultimately $Y_n$, is dependent on six parameters, namely, energy $E$, potential $V_B$, effective masses $\mu_B$ and $\mu_W$ for barriers and wells, respectively, and barrier thickness $L_B$ and well thickness $L_W$. Here, we vary the energy $E$ in the parameter $H = H(E, V_B, \mu_B, \mu_W, L_B, L_W)$ and use the membrane and stromal space thickness from the study of Daum et al [19] as the barrier thickness or length $L_B$ and well thickness or length $L_W$ or vice versa whichever has the higher potential.

## 4. Results and Discussion

The thylakoid membrane is viewed in analogy to a dielectric material in a capacitor so that the potential $V_B$ is just the energy stored in a capacitor. If we consider the membrane to be the cylindrical disk having a diameter between 0.3 $\mu$m to 0.6 $\mu$m [22] with a membrane capacitance of 1 $\mu$F cm$^{-2}$ [2], at 11 mV, the potential would be between 0.26 eV to 1.06 eV. This is the basal potential of the membrane, i.e., the resting potential of the membrane when there is no light [2]. Since wavelengths 680 nm and 700 nm carry energy at 1.82 eV and 1.77 eV, respectively, then during transition from darkness to illumination, the incident photon passes through the membranes easily.

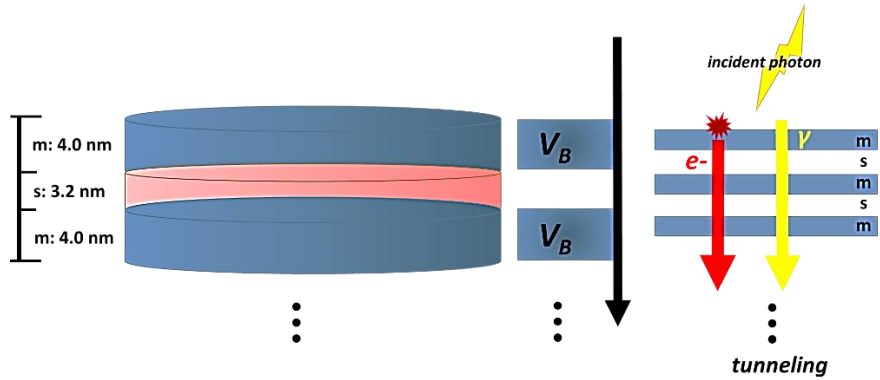

**Figure 4.** The cylindrical disk model of the thylakoid membranes: m – membrane, s – stromal gap. Photon tunnels through the membranes. The red line on the right panel indicates an electron tunneling. The yellow line indicates a photon tunneling.

Light-induced electric potential difference can peak to at least 100 mV depending on ionic concentrations and conditions, among others [2]. In our numerical simulation, the barrier potential was set at 10 eV. This gives a potential difference between 33.8 mV to 67.6 mV. From the results of Daum et al [19], the length of the barrier potential, $L_B$, is 4.0 nm while the length of the well potential, $L_W$, is 3.2 nm (see, Figure 4). Moreover, we take the two lengths to be fixed in our simulations circumventing the possibility that the grana stacks can vary vertically based on light conditions, i.e., expands at low-light conditions and contracts at high-light conditions in several studies reported in Daum et al [19]. It was also observed that plants at low-light condition may contain up to 100 grana in a stack compared to plants at high-light condition which have fewer, about 5 to 15 grana [17]. The transmission coefficients corresponding to the energy of the incident photon were calculated using these parameters and for n = 2, which is simply the basic composition of the structure, for n = 5 and for n = 10 based on the number of stacked thylakoids according to existing studies [17, 19, 22]. Figure 5 shows the simulation results. Blue graphs on the left side are spectra for when the peak is at 1.77 eV or 700 nm while the red graphs on the right side are spectra for when the peak is at 1.82 eV or 680 nm. The peaks are the incident energy whose transmission coefficient equals one. This is the condition at which resonant tunneling can occur. Increasing the number of barriers only increases the number of peaks to *n – 1* and peaking around values that are the same for n = 2. Photosynthetic organisms at low-light conditions may then compensate for their energy requirement by having a relatively greater number of stacked thylakoids where resonant tunneling can still occur. Note that there are several energies that spike up for a system with more than two barriers.

For n = 2, 5, 10, when the energy is at 1.77eV, the effective masses are, $\mu_B = \mu_W = 0.010855\mu_0$, and when the energy is at 1.82 eV, the effective masses are, $\mu_B = \mu_W = 0.010420\mu_0$ where $\mu_0$ is the free electron mass. Throughout the simulation, the effective masses are held constant except for variation only at different energies. Moreover, it is observed that the translational movement of peaks along the energy axis is very sensitive to minute changes in effective mass and the barrier potential. This may suggest that there are many combinations that would allow resonant tunneling to occur. Interestingly, a low effective mass may suggest two possibilities, electron tunneling and photon tunneling, or perhaps, both occurrences may take place. In the right panel of Figure 4, the red line shows an electron tunneling through the system after getting knocked off by incident light while the yellow line is the incident light tunneling through system. A more recent study has shown theoretically that extreme anisotropy of effective mass and low intrinsic resistance to movement, even having a zero effective mass, may be exhibited by semiconductor superlattices in a direction of electron motion. It is also suggested that electron flow can super-collimate when effective mass is low. The consequences are quite profound: higher conductivity and ultra-fast and extremely strong electron response for faster electronics and functional materials [23]. In the case of photon tunneling, different sets of parameters such as refractive index, angular frequency and angle of incidence have to be considered aside from incident energy, effective mass and potential barrier in describing a system [15]. Fortuitously, recent studies have shown the relation between angular frequency and refractive index to the effective mass of photons within a medium [24, 25, 26]. Although not an ideal comparison, the effective masses determined earlier have similar orders of magnitude [24]. Hence, having low and largely different values for effective mass is not at all impossible.

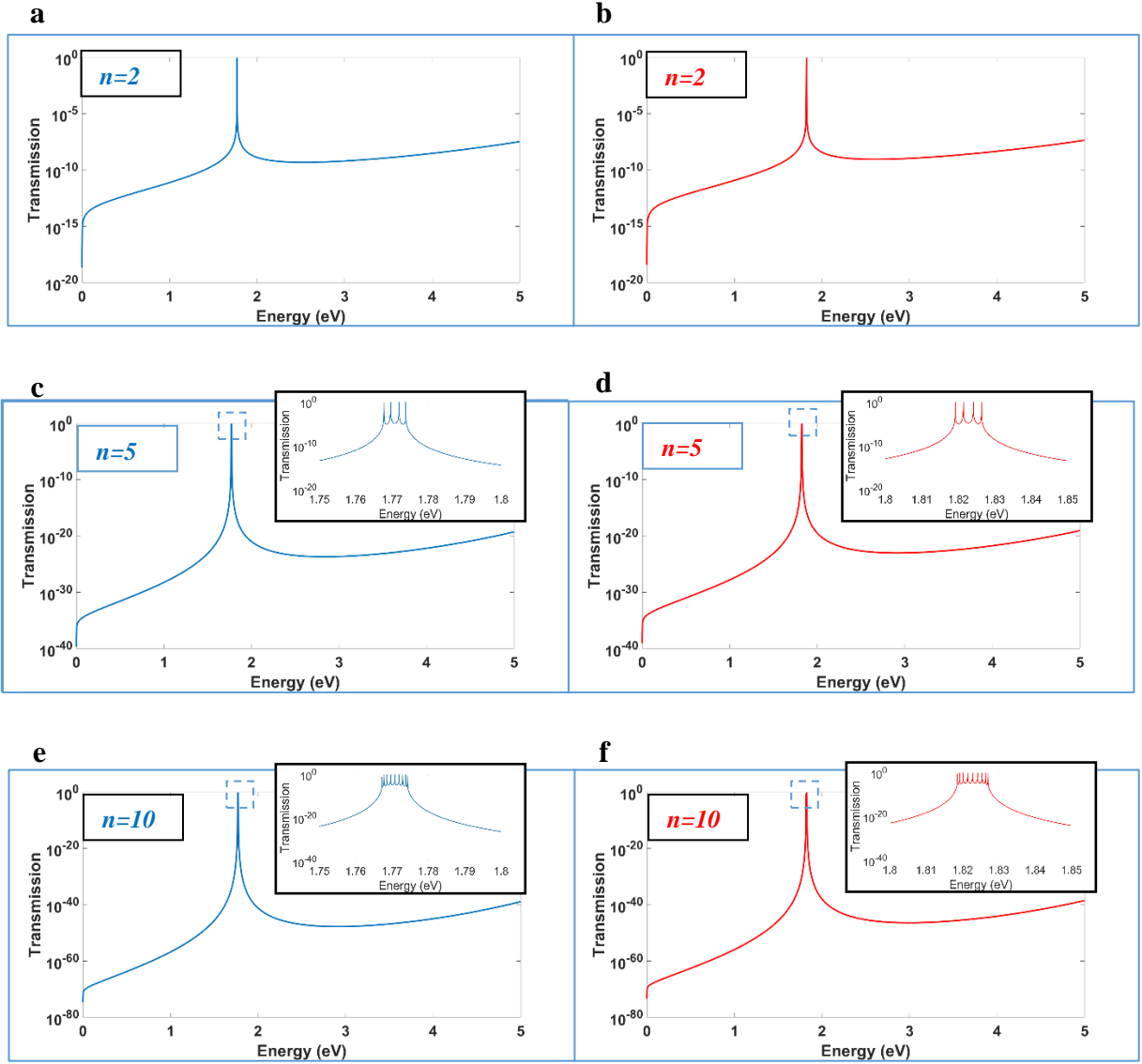

**Figure 5.** Transmission coefficient vs. the energy of incident photon. The particle tunnels through several barriers n = 2, 5, 10 with 10-eV barrier potential. (a), (c) and (e) are graphs (*blue*) that peak at around 1.77 eV or 700-nm wavelength. (b), (d) and (f) are graphs (*red*) that peak at around 1.82 eV or 680 nm.

## 5. Conclusion

For the first time, the resonant tunneling phenomenon, first theorized by Tsu and Esaki [10] and since then has been proven experimentally and widely applied in semiconductor systems, is applied in a natural photosynthetic system. We showed the possible occurrence of resonant tunneling by working with the geometrical dimensions and empirical measurements of the stacked structure of grana thylakoids and considering them as superlattices. This behavior originally observed in semiconductors may contribute to the internal quantum efficiency or photon-energy-charge energy conversion in photosynthesis because the energy gets transported across the whole grana stacks with minimal reduction. In all of our simulations, the transmission coefficients of the incident energies would peak, i.e., transmission coefficient $T_n = 1$ for $n$ barriers, at energies widely known to be used by plants corresponding to wavelengths of 680 nm and 700nm. This is the condition stipulated for resonant tunneling. Such analysis also addresses the problem of deep thylakoid membranes not readily accessed by light because of their location near or at the bottom of 20 or more stacked thylakoids. Indeed, the number of barriers does not even matter since their respective transmission spectra can peak at the said energies. We note that in our simulation, a choice is made in fixing the lengths of the barrier and potential wells since different specimens may have differences in dimensions [19]. In addition, the light-induced potential difference across the membrane varies significantly depending on the conditions [2, 27]. However, our simulation reveals that a minor tweaking of parameters causes significant changes to transmission spectra. This allows for many configurations or conditions where resonant tunneling can occur.

Another feature for resonance tunneling in photosynthetic systems is a very low effective mass. The low effective masses allow tunneling particles to interact readily and move easily across the membranes. Two types of particles could tunnel through grana thylakoids to achieve a supply of excited electrons even at the bottom of a grana stack hardly accessible to light for photosynthesis to occur at thylakoid membranes. The first type is an electron with low effective mass arising from interactions with the medium. For instance, a transport regime in a semiconductor superlattice with low effective mass for an electron has been discussed [23]. In fact, probing the low effective masses of tunneling electrons has been a continuing experimentally active area [28-30]. On the other hand, a low but finite effective mass for the photon has also been crucial in the creation of a stable luminous quantum fluid of light when photons are spatially confined [24-26]. Further experimentation would be required to determine whether it is the photon or electron that tunnels, or both, which could account for the high internal quantum efficiency of photosynthesis. It is vital to explore this further as resonant tunneling in photosynthesis could have implications in the design of more efficient solar cells.

**ACKNOWLEDGMENTS**

K. M. G. gratefully acknowledges the support of the Department of Science and Technology - Science Education Institute through the Accelerated Science and Technology Human Resource Development Program – National Science Consortium (ASTHRDP-NSC).

**DECLARATIONS**

The authors have no conflicts of interest to declare that are relevant to the content of this article.